# Aadhaar Card: Challenges and Impact on Digital Transformation

Raja Siddharth Raju[1], Sukhdev Singh[1,2], Kiran Khatter[1,2]

[1]Department of Computer Science and Engineering
Manav Rachna International University, Faridabad, Haryana-121004, India.
[2]Accendere Knowledge Management Services Pvt. Ltd. Chennai-600101, India.

rsrajuofficial@gmail.com,
sukhdev200@gmail.com
kirankhatter@gmail.com

**Abstract**

**Objectives**: This paper presents a brief review on Aadhaar card, and discusses the scope and advantages of linking Aadhaar card to various systems. Further we present various cases in which Aadhaar card may pose security threats. The observations of Supreme Court of India are also presented in this paper followed by a discussion on the loopholes in the existing system.

**Methods:** We conducted literature survey based on the various research articles, leading newspapers, case studies and the observations of Supreme Court of India, and categorized the various cases into three categories.

**Findings:** Aadhaar project is one of the significant projects in India to bring the universal trend of digital innovation. The launch of this project was focused on the inter-operability of various e-governance functionalities to ensure the optimal utilization of Information, Communication and Technology Infrastructure. Towards this Government of India has recently made Aadhaar card mandatory for many government applications, and also has promoted Aadhaar enabled transactions.

**Improvements:** There are many issues related to security and privacy of the Aadhaar data need to be addressed. This paper highlights such cases.

1. Introduction

Aadhaar project was introduced under the scheme 'UIDAI' (**U**nique **I**dentification **A**uthority of **I**ndia) by the UPA (**U**nited **P**rogressive **A**lliance) government in year 2009. In fact in 1999, Former Prime Minister of India Shri Atal Biharee Vajpayee had suggested identity cards for the people living in the border area, and the idea was later accepted in 2001 by the Former Home Minister Lal Krishna Advani, who recommended a multi-purpose National Identity Card. Later in 2009, UIDAI came into existence under the UPA government, and Nandan Nilekani, co-founder of Infosys was appointed as the chairman of the Aadhaar Project. Aadhaar card contains the demographic features such as name of the citizen, Father/Mother's name, Date of Birth, Sex, address of the citizen, and biometric features such as photograph, fingerprints and iris (eye) details. The demographic features as well as in the form of **Q**uick **R**esponse (QR) code along with a 12-digit unique identity number called, Aadhaar, are printed on the card issued to every citizen. All the demographic and biometric data are stored into one centralized database, and this project has been reported as a world's largest database management and Biometric ID system respectively by Forbes [1] and The Times of India [2]. The UIDAI project provides the online support to change the demographic data of Aadhaar Card using SSUP (**S**elf **S**ervice **U**pdate **P**ortal) from UIDAI official website (uidai.gov.in). For an instance to change the name, one needs to submit the Gazette Notification of India mentioning that 'required person's name has been changed from old name to new name'. To update DOB (**D**ate **o**f **B**irth), the required documents are Birth Certificate issued by the District Municipal Corporation, and for the people who don't have a birth certificate and were born before 1989, they can provide an affidavit to change their DOB. Further, if one don't have the required document to change the DOB, then SSLC (**S**econdary **S**chool **L**eaving **C**ertificate) or Passport can also be taken into consideration. For changes in address, electricity bill, landline bill, credit card bill less than three months old, bank passbook, Voter ID, Passport or a rental agreement, and the scanned copy of proof of identity is sufficient. Changes can also be made to the Gender and Mobile number as well, and proof of identity is required for these purposes. For all the demographic changes the authentication is being checked through an OTP (**O**ne **T**ime **P**assword) sent to the registered mobile number. However, the biometric data can't be changed.

Nowadays the government of India is linking the Aadhaar card with many government functionalities, but there are many security and privacy issues of the Aadhaar database need to be addressed. In this paper, Section 2 discusses the scope and advantages of linking Aadhaar card to various systems. Section 3 present case studies in which the implementation of Aadhaar card may lead to security threats. In Section 4 the observations of the honourable Supreme Court of India are discussed. Further Section 5 presents the discussion on the loopholes in the existing system along with conclusion.

2. Scope of Aadhaar Card

The objective of this section is to highlight the scope and advantages of linking Aadhaar card to various systems. The government of India has been linking the Aadhaar card with various government schemes such as for cooking gas subsidies, house allotments, school scholarships, admission into remand and welfare houses, passports, e-lockers (eg. Digilocker), for archiving documents, bank accounts under

PMJDY (**P**randhan **M**antri **J**an **D**han **Y**ojana), provident funds account, pensions, driving license, insurance policies, loan waivers and many more [3]. Recently it has also been made mandatory for ATM Cash Transaction [4], railway reservation [5] and applying PAN (**P**ermanent **A**ccount **N**umber) card, and filing income tax returns [6]. In fact in 2016, Aadhaar Bill (Targeted Delivery of Financial and Other Subsidies, Benefits and Services) was introduced as a Money Bill in Lok Sabha, aimed to provide for good governance [7]. In this bill, Aadhaar card was made mandatory for authentication purposes like salary payment, pension schemes, school enrolment, train booking, for getting driving license, to get a mobile sim, to use a cyber café etc. [8]. Recent news suggest that UGC (**U**niversity **G**rants **C**ommission) instructed the universities to include a photograph and seed the 12 digit Aadhaar number on the mark sheet as well as on the certificates to bring consistency and transparency. Further with security features it would eliminate the duplication of the mark sheet [9]. Apart from all these next we present literature review suggesting the linking of Aadhaar card to various systems and its advantages. First we discuss railway reservation system.

## 2.1 Railway Reservation System

India is the world's largest human transportation in which over 20 million passengers travel daily through Indian railways. Therefore, the Government of India is taking much initiative to make railways at its best, and soon it will be witnessing the bullet trains running on the track. In the existing Indian railway reservation system, passengers can book the tickets online through the www.irctc.co.in, a site managed by IRCTC (**I**ndian **R**ailway **C**atering and **T**ourism **C**orporation **L**td.). In recent years IRCTC has improved a lot both in terms of the efficiency and blocking fraudulent bulk ticket bookings. For an instance online booking supports an auto fill form in which the user simply needs to login at IRCTC website and the already filled details are transferred to online booking form. It may help the people but it had been mostly used by touts for their financial gain unless the Capthcha was introduced. It has also been observed that the IRCTC site consists of fake ID's created by impersonation. So prevent cases of impersonation, fraudulent booking and from blocking bulk tickets at once, IRCTC is going to make Aadhaar mandatory for online railway ticket booking [5]. In fact from April 1, 2017, Aadhaar Number would be required for one-time registration in the IRCTC ticketing websites for which a three-month trial is already going on. It is to be noticed that in the existing system only identity card is checked while travelling which is not enough, and it may lead to various terrorist attacks like 26/11, Samjhauta Express etc. To enhance the security linking of Aadhaar card may be very useful. For an instance in their work [10] proposed Aadhaar Card based reservation system in which the Aadhaar card will be mandatory to book the tickets, and further the passenger will go through two stages to enter into the platform. The first stage is the Security Check in which passenger will be authenticated by his Aadhaar Card, and the biometric data of the person will be matched with the Aadhaar Database in the second stage. After successfully going through both stages the passenger will get an update at his/her mobile number with his ticket that whether he can enter the specified platform. [11] also proposed a model which overcomes the drawback of online ticket reservation. In their proposal if one person wants to purchase multiple tickets, then he first has to authenticate himself with his Aadhaar card and an OTP (**O**ne **T**ime **P**assword) sent to his/her registered mobile number, and further Aadhaar card number of all the passengers will be required. Once the ticket is successfully booked, all the passenger would be notified so any false/fake ticket booking can be observed easily. Therefore the linking of Aadhaar card can help a lot to improve the existing railway reservation system.

**2.2 ATM Security**

According to Indian Express [12] there are over 2,15,039 number of ATM (**A**utomated **T**eller **M**achine) till the end of June 2016 in India. Further in August 2016, over 75.6 core transaction took place in which 8 crore were through cheque clearance, 7 crore through mobile banking and rest were through ATM. It shows the importance of the cash transactions through ATMs. However in recent past various ATM fraudulent cases have been observed. One of the major breach of ATM security is found through skimmer device in which the culprit put a key logger under the components of the (like keypad or magnetic tape reader) ATM, and when a user insert the card and pin, it use to store the credential information [13]. Another ATM security breach is through Key Jamming, in which the keys that are 'Enter/OK' or 'Cancel' are being jammed. So after entering the PIN (**P**ersonal **I**dentification **N**umber), jammed button leads to hijacking of the session [14]. To enhance the security of the ATMs [15] proposed the model of Biometric ATM enabled with the fingerprint approach to identification. Authors suggested the use of Fingerprint for authentication purpose apart from PIN verification. However various physical factors such as need of high definitions camera in ATM, extreme weather conditions some are barrier to implement Biometric enabled ATM. Recently [16] suggested ATM transaction using Aadhaar card system and OTP. In this system the Aadhaar Card details will be mandatory to access the account with OTP. The main idea for Aadhaar card enabled ATM is that it consists of various biometric details which substitutes same to biometric enabled ATM. In fact in June 2016 DCB (**D**evelopment **C**ooperative **B**ank) has already launched its first Aadhaar enabled ATM at Bengaluru, Karnataka. It is the first ATM in India which uses Aadhaar number and Aadhaar Biometrics to withdraw cash and other transaction [17]. This service has major benefits like no need to carry ATM, blocking ATM card in case of lost, and moreover biometrics are more secure than the ATM pin. Therefore in future Aadhaar card can be seen as a medium for Banking.

**2.3 Cloud Based E-Voting**

In India currently the voting is done through the EVM (**E**lectronic **V**oting **M**achine). EVM comprises of 2 machines that are Ballot Unit in which the voter pushes the button to vote for the party, and a Control Unit which is accessed by the poll workers to count the votes. The major drawbacks of the existing system are high cost and high man power. Further one of the major frauds is booth capturing in which the party locals may capture the booth for gaining majority of the votes for a political party. Recently, in the

UP (**U**ttar **P**radesh) election 2017, one of the opposition party's leader, Shrimati Mayawati has also raised the issue of tampering the EVM's [18]. A few cases in which the faulty EVM's favouring only the ruling BJP party have also been observed. In existing system it has also been observed that the most of polling remains 60-65% because the public have no interest in polling, and further if they are interested sometimes they may not be in the position to cast their vote. To improve the system cloud based E-Voting can be seen as a future polling system. [19] suggested the Aadhaar based model to eliminate the drawbacks of existing voting system. Authors discussed the cloud based server connected with two parts. At first part, the Voter/User is connected at the one end of the cloud server, and he/she will access the voting technology through cloud computing. However he cannot have the direct access to the voting line until and unless he/she has been authenticated. In the second part, it is concerned with Aadhaar database in which the 1$^{st}$ server would be connected with the cloud server. The user that would try to gain access to vote has to be authenticated at initial stage by sending his scanned copy of iris and fingerprint to the Aadhaar database, and when the identity is authenticated the Aadhaar server will generate a e-ballot paper on which the user can fill the voting details. On the 2$^{nd}$ server connected with the cloud server i.e. ECS (**E**lection **C**ommission **S**erver), when the E-Ballot paper has been submitted, the paper would be encrypted with a small algorithm and it is being sent to the ECS and then it is stored in the Election Commission Database. At last, when the voting is done, the token would be generated successfully to ensure that the vote has been counted in. In future Aadhaar card can be used in India for e-voting.

**2.4 Aadhaar e-KYC Services**

KYC (**K**now **Y**our **C**ustomer) is generally a form which verifies the identity of its clients. In KYC, a person has to fill his demographic details and to provide the documents in support of the given details. The major drawback of KYC is that it exploits to document forgeries, requires more human effort, human interference, and less availability etc. Recently, UIDAI have launched a service "Aadhaar e-KYC (**E**lectronic-**K**now **Y**our **C**ustomer) eService" in which KYC would be automatically filled with the details of Aadhaar card. The main objective to implement this service is that it offers biometric based validation which eliminates all the extra machines required to record biometric details of a person. The I-T (**I**ncome **T**ax) Department is also considering to issue of PAN (**P**ermanent **A**ccount **N**umber) card on the basis of Aadhaar e-KYC facility [20]. In fact NSDL (**N**ational **S**ecurities **D**epositor **L**imited) and UTIITIL (**UTI I**nfrastructure **T**echnology and **S**ervice **L**imited) provide Aadhaar e-KYC Services. Recently, the Government of India has declared that the Aadhaar card must be linked with the PAN card till 31 December, 2018 and PAN card would be mandatory for hotel bills above Rs.50,000 and jewellery bought for above Rs. 2,00,000. Also for tax payers, it is mandatory to link Aadhaar card with PAN card before July 1, 2018 else, their filings would be rejected. The article of Roy [21] gives a brief review on linking banking details with demat accounts to form a single account. Author suggest that this idea can be stretched to basically cover field of financial assets like bank details, mutual fund, insurance, provident fund, pension fund account, and demat account. These financial assets of an individual can be combined into a one single unit with the linkage to Aadhaar number. Linking Aadhaar to all financial assets in different banks will help in ease access of the banking details [21].

**2.5 Denture Identification**

Denture Identification is a method of recognizing an individual who has been affected in course of a natural disaster. It includes a unique number to be placed in the mould area of mouth. Various countries have adopted this method, like in USA 21 states have been made mandatory for denture marking with their **SSN** (**S**ocial **S**ecurity **N**umber). In Australia, tax file numbers are used for Denture Marking and in Sweden, unique personal identification is used for denture marking. Recently in India in June 2013, a natural disaster took place in Uttarakhand in which many dead bodies swept away, and there were several dead bodies who were not been able to recognize. In such situations, Denture marking may help. [22] proposed a model to integrate the Aadhaar number with the dental labelling. Authors suggested the printing of Aadhaar number on a paper laminated with the thermoplastic sheet on both sides to place into the mould space of polish surface side with additional layer of heat polymerized clear acrylic resin. The proposed methodology has advantages such as simplicity in performance, availability of materials, and enhancement of identity management, inexpensive and naturally sluggish after being placed in the denture. However, some of the drawbacks of the proposed method are that the acrylic resin does not survive temperatures beyond 300 degrees Celsius, and information cannot be retrieved if the denture fractures take place in the area of denture labelling. Authors have further suggested that in such incidences, mandible lingual flange is a safe location relatively.

**2.6 E-health Care**

In India healthcare is the primary responsibility as it have less infrastructure and lack of doctors in rural areas. According to the statistics, 75% of doctor's work in urban sectors, 23% works in semi urban sector and only 2% work in rural sector [23]. To provide the services to both urban and rural areas, E-health care was introduced in India using Internet Technology to eliminate the possible threats due to insiders. For an instance, insiders who make guilty mistakes and cause disclosure of confidential information of the person, who knowingly access information for profit, who gains access to information for revenge against outsiders or employees. Apollo Healthcare, ISRO, and CSIR provide E-healthcare services in India. To overcome the possible threats due to insiders, [24] proposed the authentication and authorization model of E-health care using Aadhaar card. The suggested model contains two phases. The first phase is authorization in which the user is identified for role based authorization like patient, specialist, nurse etc., and based on the role privileges are granted. For an instance if a patient is suffering from heart disease, then his personal details can only be accessed by the specialist. The second phase is authentication in which the user identity is verified of a role in e-health care service using Aadhaar card. In the suggested Aadhaar based E-health system, a user has to first register with the e-health service system through administrative agent who will check the role of the user. If the role is authorized i.e. if the role is either GP or Patient or nurse or specialist, then only original authorization reference will be generated (user cannot access e-healthcare system). After this, the user will get the temporary authorization reference number where the user is asked to provide the Aadhaar number. The user is authenticated online at CIDR (**C**entre **I**dentification **R**epository) maintained at UIDAI. If the authentication is successful formal authorization reference will be generated. If it is normal the user is verified for the different privileges (reading/writing) into the e-health service System. If the status of the user is satisfied according to the authorization policy, then he will have the permission to access the relevant record. Moreover if the status of the user is not normal like in emergency, multiple authentication and authorization would be compulsory. The general physician is required to authenticate through Emergency Identification number provided from the office of Chief Medical Officer. If the privilege is authorized, permission is given for accessing the e-healthcare system.

## 2.7 Municipal Corporations

In India Municipal Corporations are local government bodies that work towards the development of cities having more than 10 lakh citizens. They are responsible for managing community services like public healthcare, sewage, electricity, road etc. Though, the delivering of the services by the municipal corporation is bad. Complaints registered in Municipal Corporation sometimes have zero response and public health is put in danger as lack of functionality. [25] proposed a model for effective solution towards complaints registered in Municipal Corporation using Aadhaar card. In the proposed model, Aadhaar card is used for the authentication purpose as before the registering a complaint, one must register himself on the municipal corporation using Aadhaar number. Once entered, they will get a message from the website which will have login credentials. Once user logs in, then he/she have to select a department in which he/she want to register a complaint. After that, problems from the given list can be selected or if problem is not in the list, then the message can be dropped. After successful complaint registration, complaint is saved in the municipal database with date, time and user details with the help of Aadhaar card. On the other hand, in Municipal Corporation, each department has an in charge that is being provided with a login credentials provided by the municipal corporation. When the in charge logins, he is able to see the complaints registered under his department and then the work is allotted according to priorities. Also, they are responsible to check that work is done by the people under them. To avoid the misuse of the system, a black mark is given to each bogus request generated by the user. Limit of black mark is 3 and if it exceeds, then the user is not viable to make a request again. Priority is given on the basis of date, time and number of people who registered the same type of complaint. Users can also check the status of their complaint. In fact there are four stages of the status, when you register a complaint; the status is pending which is first and foremost. After pending, it will move to the status progress which states that the work on complaint has started. If the status is either in progress or in pending after exceeding the default time allocated, then the status shifts to 'Forwarded to Higher Officials' where higher official takes care of it. If the problem still is unresolved after 24 hours, then the status changes to 'Forwarded to Press' which cannot be reverted back to completed status. Once the work is complete, then the status changes to 'Completed'. When multiple people register complaint for same problem, then employees must ensure that at least one amongst them status changes to completed. The idea to propose this model was to establish a direct communication link between municipal body and the citizen. It may help in bringing transparency in the system and efficiency towards the municipality.

## 2.8 PDS (Public Distribution System)

PDS (**P**ublic **D**istribution **S**ystem) was implemented in India in 1965 for the poor people who can have the food delivered at a low cost or free of cost. In fact the government bought the food grains from the farmer at a procurement price and sell the food grains through the PDS. In 1997, the universal PDS system (where each and every one was eligible, even rich people can opt for PDS for low cost food grains) was abolished, and the entire country was divided into two parts, APL (**A**bove **P**overty **L**ine) and BPL (**B**elow **P**overty **L**ine). Now, only BPL citizens are only allowed to have an access to the PDS at low or free of cost [26]. Recently the government has linked Aadhaar card with the PDS system to overcome the frauds. The major reason to implement Aadhaar based PDS system is that there were increase in the number of 'Rice Mafia', people who just use the rice from the government using fake ID's and sell it outside to someone else for profit [27]. It is to be noticed that in previous PDS, each citizen with ration card were allotted with specific FPS (**F**air **P**rice **S**hop) or ration shops. However linking of Aadhaar card provides the access of any nearby FSP's or Ration Shops.

## 2.9 Aadhaar Pay

In India demonetization took place on 8 November, 2016 but it didn't suit the daily lifestyle of the people. Many people had to queue up to exchange money and withdraw the new currency from the bank and ATM. Therefore, this made cashless transaction come into existence in India. There were various merchants who started accepting 'Paytm'. Aadhaar card cashless transaction also started in which the user has to simply enter their Aadhaar card number and the amount would be debited from their account automatically. In fact, IDFC Bank Ltd becomes the first Indian Bank to launch Aadhaar Pay [28]. IDFC bank launches its first biometric enabled transaction method in which the person can pay the merchants using his/her fingerprint. The idea was implemented for the people who don't have smart phone or if owning one, but without banking features. IDFC bank distributes a biometric enabled device to the merchants to connect it to smart phone for transaction purposes. Biometric enabled transaction or Aadhaar pay would soon be launched by four more banks that are Syndicate Bank, Punjab National Bank, Bank of Baroda and Inducing Bank. Also, IDFC bank has some concessions for the merchants who are willing to take part in this mission as 0.25 percent of every transaction is credited to the merchants account as a promotional offer.

## 2.10 Digilocker

In India, it is very much difficult to maintain the papered documents and achieving it from the government offices. The major drawbacks of the papered system are lack of transparency, high cost, requires human effort, and time consumption etc. Recently in February 2016 Government of India introduced 'Dig locker', a cloud based personal space for storing documents. To sign up for Dig locker, one must have an Aadhaar Card. It provides the personal electronic space up to 10 MB for storing documents. This service enables the storing of the documents issued by the Government Departments as well as your University or High School certificates, URI (**U**niform **R**esource **I**ndicator) of each document issued by issuer departments. The linking of Aadhaar card gives a benefit of e-Signing the documents. One of the main advantages of Dig locker is automatically verified document by the specific Government Agencies. The Government is also planning to extend its storage space up to 1GB.

## 2.11 MGNREGA

MGNREGA (**M**ahatma **G**andhi **N**ational **R**ural **E**mployment **G**uarantee **A**ct) earlier known as NREGA (**N**ational **R**ural **E**mployment **G**uarantee **A**ct) is an Indian labour law or social security measures in which it provides at least 100 days of wage employment in a financial year to every household whose adult member does some unskilled manual work so as to enhance the livelihood of the households in rural areas [29]. Recently, Indian Government has subsequently made mandatory to have an Aadhaar card to apply for MGNREGA benefits. The main aim to link Aadhaar number with MGNREGA was to bring transparency and efficiency in the welfare schemes. Further it ensures that the money is being transferred to the verified bank account using DBT (**D**irect **B**enefit **T**ransfer). One more advantage is to keep track of a beneficiary with his information including biometric scans [29]. It is to be noticed that according to the Supreme Court of India order 2015, Aadhaar is not mandatory for every welfare scheme except cooking gas subsidy and PDS system. But in 2016, Aadhaar extended the welfare schemes to MGNREGA and EPS (**E**mployee **P**ension **S**cheme) [30].

## 2.12 National Pension Scheme

PFRDA (**P**ension **F**und **R**egulatory and **D**evelopment **A**uthority) was established in 2003 to develop and regulate pension sectors in India. In 2004, NPS (**N**ational **P**ension **S**cheme) was established to provide pension to citizens all over India. Recently NPS has been restructured and the pensioners are asked to link Aadhaar number with pension scheme. The main objective of linking Aadhaar is to prevent the delaying of the payments every month [31].

## 2.13 Pardhan Mantri Jan Dhan Yojna

PMJDY (**P**radhan **M**antri **J**an **D**han **Y**ojana) is a National Mission for financial inclusion in which it allows a citizen to open a savings account with no minimum balance i.e. zero balance account. The benefits of this scheme include the interest on deposits, accidental insurance up to Rs. 1 Lakh, overdraft facility up to Rs. 5000, transfer money within India, benefit of government scheme will get DBT, access to pension and insurance, and free debit card (Rupay Card) to withdraw money. Recently, Aadhaar has been linked with PMJDY. Therefore, a user can open an account with an Aadhaar card as a proof, and can avail the government welfare schemes benefits plus to avoid the misuse of the overdraft facility given to the user [32].

### 2.14 Aadhaar Mandatory to Fly on a Plane

According to the report by Times of India [33], the government has asked an IT company, Wipro to make a blueprint for Aadhaar based biometric access to flyers on all airports in India. The Ministry of Civil Aviation is thinking of linking Aadhaar with all flyers booking tickets. The idea is that when a passenger is booking tickets, he/she must provide his/her Aadhaar number. Once a passenger reaches the airport, at every touch point he has to place his thumb. Similar process can be iterated at the time of check-in. However, the major drawback is for the tourist who doesn't hold an Aadhaar card.

### 2.15 Maps of India

Recently, Government of India had launched an official website from which public can download maps of India, but Aadhaar is made mandatory. Basically, SGI (**S**urvey **G**eneral of **I**ndia) completed its 250 years, so they introduced a web portal from which public can download 3000 maps prepared by SGI but public is limited to 3 maps per day. Though, they have 5000 maps from which 3000 are made available, 1700 are to introduced soon and rest are awaiting clearance. The main reason to implement Aadhaar linkage was to avail it only to the Indian citizens [34].

## 3 Case Studies

In the previous section we discussed the advantages of linking Aadhar card to various systems, but at the same time Aadhaar project have also raised many questions. The objective of this section is to discuss the issues related to data privacy and information loss, implementation loopholes, and the issues due to lack of awareness. Next we discuss issues of data privacy and information loss.

### 3.1 Data Privacy and Information Loss

Data is an asset of an organization, and Privacy is some sort of assurance that an individual requires from an organization. Therefore Data privacy together refers to the ability of an organization that determines which data has to be shared with third party. As the Aadhaar card contains both the demographic and biometric data, so it becomes a risk for an individual as well as to the government if the data are insecure. It is to be noticed that Clause 30 of IT Act 2000 states that biometric or demographic data are recognized as an 'electronic and sensitive data of an individual', and if someone tries to steal it, there is a Clause 34-

47 under Chapter VII of IT Act 2000 which deals with punishment related to it, and also is entitled as 'Offences and Penalties' [35].

Though there are strict laws but still whether the data in Aadhaar database are secure or not has always been a question. According to The Times of India [36], Maharashtra accepted that their 3 lakhs of Aadhaar data got lost with PAN. The incident happened when the IT Department were uploading the biometric information and PAN data to the UIDAI centralized server that is in Bengaluru (then Bangalore) from Mumbai, due to the crash of hard disk. In fact the data were being uploaded and encrypted using strong algorithm, and when the Headquarters were downloading the data, they couldn't decrypt it. Therefore many applicants, who complained about this, were asked to re-register for it. Later the State (Mumbai) IT department stated that the data belonged to people of Mumbai, and the lost data are being fully secured which can only be opened if you have 'keys and multi clues'. The State ensures that the data are safe but such type of issues has already raised serious concern. In a recent case, Sakshi Dhoni, wife of Indian cricketer MS Dhoni tweeted to the Union of Law & Justice plus the Ministry of Electronics and IT about the Aadhaar data of MS Dhoni being leaked by the CSC e-Governance Services India Ltd [37]. The CSC e-Governance Services India Ltd. had posted a photo of MS Dhoni fingerprint being scanned as well as the screenshot of the Aadhaar data of MS Dhoni. The shocking thing was that the Electronics and IT Minister also liked the tweet and retweeted the photo of MS Dhoni's fingerprint being scanned by the CSC agency. Later the UIDAI took the strong step and blacklisted the CSC e-Governance Services India Ltd for next 10 years. In spite of all these rule and regulations sharing the information from a partner company raise the issue that whether any privacy is left and does this ensure that whether Aadhaar data is in the right hands or not. According to the sources of Indian Express [38], recently first time the NDA Government has admitted that the Aadhaar data had been leaked to the public domain. However the government had been ignoring the fact that Aadhaar is a sensitive data and assuring us by saying that Aadhaar is fully secured and it can't be breached easily. As the Aadhaar project has the largest database management the information loss or security breach to Aadhaar database can be a serious threat for India.

Again in a shocking incident, the information consisting names, addresses, Aadhaar numbers and bank accounts of more than a million beneficiaries of Jharkhand's old age pension scheme, have been compromised by a programming error on a website maintained by the Jharkhand Directorate of Social Security [39]. It is to be noticed that the Jharkhand government has over 1.6 million pensioners, 1.4 million of whom have seeded their bank accounts with their Aadhaar numbers to avail of direct bank transfers for their monthly pensions. As the information of all the citizens can be accessed freely by logging onto the website, this case raise the serious threats of linking Aadhaar card to various government functionalities.

In a similar case in Kerala, Aadhaar data of over 35 lakhs of pensioner has been leaked from the Kerala state pension department. All those 35 lakhs of pensioners had linked their Aadhaar number and bank account as required by the "direct benefit transfer" scheme. The service pension website had put up their names, addresses, phone numbers, bank account numbers, Aadhaar numbers and photographs for anyone to download in stark violation of the Aadhaar Act. Furthermore the site also had the pension id used to draw information, and the data has been pulled from the website only after the news created a stir [40].

In Chandigarh, food and supplies and consumer affairs department shared the UID numbers of number of people on their website. It was said that even ration cards of the person, Date of Birth, spouse name details were displayed on their public domain. In fact Ministry of Water and Sanitation, which is considered as one arm of Swatch Bharat Mission too publicized the Aadhaar details of the citizens with details like Voter ID number, ration card number and their caste status [41]. However due to various cases of data leaks from government domains, the central government has recently circulated a set of 27 do's and 9 don'ts on data handling and instructions to encrypt sensitive data with a legal consequences. Further each department has been asked to review their public domain to check if there is any personal data on display, and to allot an official who must be responsible for Aadhaar data protection [42].

### 3.2 Implementation Loopholes

Government of India has been linking each and every welfare services and benefits to Aadhaar card. But linking is not sufficient, and from the day Aadhaar project has started it had been in the news for several implementation loopholes. There are various cases occurred in which the Aadhaar implementations had faced problems. The objective of this section is to discuss such cases.

According to The Times of India [43], there was an Aadhaar controversy in which the Aadhaar card were being considered invalid on the various factors. In this case a senior citizen got his Aadhaar card without any hassle or without any problem, but the problem aroused when he got the Aadhaar card mentioning the '**Year of Birth**' instead of '**Date of Birth**' which was considered as an invalid Aadhaar card. Later the Secretary of State (Mumbai) IT Department considered it to be valid as the senior citizens who were born before the year 1989, can use Year of Birth as they didn't have the provision for birth certificate at that time. Recently, Aadhaar has been made mandatory to be linked with PAN card, since then various cases of mismatching names on PAN card and Aadhaar card have also been reported [44]. The main reason is that Aadhaar does not require to disclose the name of the citizen without initials where as the PAN requires the disclosure of full name with the initials. Due to this, many people are not able to link PAN with Aadhaar card. As per the finance act 2017, now Aadhaar is mandatory for applying a fresh PAN application and for filing Income Tax returns. Further government is also saying that the existing PAN would be cancelled if Aadhaar has not been linked with it; the reason is to control tax evasion and eliminate multiple PAN's. Therefore the linkage of Aadhaar to PAN is a good initiative but name of Aadhaar card has not paid much attention while implementing this project. Though such types of problems are occurring, it is to be noticed that Income Tax department made mandatory to link Aadhaar card with bank accounts by 30 April 2017 to self-certify them to comply with FACTA (**F**oreign **T**ax **C**ompliance **A**ct) regulations [44].

According to Live mint [45], UIDAI filed a complaint on which Delhi police has lodged an FIR in which two different names enrolled with same biometric. The Deputy Director of UIDAI regional office in Pragati Maiden, Delhi told police that on March 18, a person named Raj Kishore Roy enrolled for Aadhaar and submitted his demographic and biometric details. However UIDAI found that on March 17, a person named Deben Roy enrolled for Aadhaar with same biometric information. This example also raises serious concerns. However later UIDAI lodged a complaint under Aadhar act as cheating by impersonation.

According to the report by Times of India [46], UIDAI filed a complaint with Delhi Police against Axis Bank Ltd., business correspondents Suvidhaa Info serve and ensign provider, eMudhra stating that they have attempted unauthorized identity theft by means of illegally storing of Aadhaar biometrics. It was found that 397 transactions were made with the same biometric between July 14, 2016 to February 19, 2017 in which 194 transactions were from Axis Bank, 112 from eMudhra and 91 through Suvidhaa Infoserve. The problem was detected when multiple transactions took place using the same biometric. The Axis Bank spokesperson told to livemint.com that it was the developer Suvidhaa, who carried out live-based Aadhaar Authentication and that goes against the government that claims that Aadhaar is not penetrable. UIDAI gave them the time till 27$^{th}$ of February 2017 to explain why they did this. However UIDAI have banned the Aadhaar enabled system temporarily for the three firms i.e. Axis Bank Ltd. e-Mudhra and Suvindhaa Infoserve. Chapter VII Clause 35 of Aadhaar act 2016 states that "Whoever, with the intention of causing harm or mischief to an Aadhaar number holder, or with the intention of appropriating the identity of an Aadhaar number holder changes or attempts to change any demographic information or biometric information of an Aadhaar number holder by impersonating or attempting to impersonate another person, dead or alive, real or imaginary, shall be punishable with imprisonment for a term which may extend to three years and shall also be liable to a fine which may extend to ten thousand rupees".

On February 2017, a FIR (**F**irst **I**nformation **R**eport) was filed against an individual known to be Sameer Kochhar, who leads Gurgaon think tank Skoch Development Foundation. The complaint was filed for his article "Is a Deep State at work to Steal Digital India?, in which he mentioned the Aadhaar security vulnerability and included a video on how the Axis Bank fraud transactions case took place [47]. The Authority Chief Executive responded it in Twitter by tweeting "The video is fake and asked Kochhar to stop spreading rumours". The DCP (**D**eputy **C**ommissioner of **P**olice) confirmed that "UIDAI has filed a police complaint against Kochhar regarding putting a fake video on an article in Google. Therefore, the a senior police official confirmed that the case filed by UIDAI was under section 37 of Aadhaar Act which states that, "Whoever, intentionally discloses, transmits, copies or otherwise disseminates any identity information collected in the course of enrolment or authentication to any person not authorised under this Act or regulations made there under or in any agreement or arrangement entered into pursuant to the provisions of this Act, shall be punishable with imprisonment for a term which may extend to three years or with a fine which may extend to ten thousand rupees or, in the case of a company, with a fine which may extend to one lakh rupees, or with both (Aadhaar Act 2016)".

**3.3 Lack of Awareness**

There are many cases where the security threats are due to the lack of awareness in the people. The objective of this section is to discuss such cases. Recently it has been observed by the UIDAI that there are various e-commerce websites who are charging to print Aadhaar data on a plastic card [48]. In fact some entities were charging Rs.50 to Rs.200 and making fool of the customers stating that Aadhaar printed on the plastic card are said to be 'Smart Card'. However, UIDAI has stated that "There is no such concept called Smart Card, and Aadhaar data printed on a normal sheet of paper is enough as a proof." It cautioned the Government as the websites may not be only interested in this scheme but the sensitive data in the hands of the e-commerce website can be misused. Therefore UIDAI warned several e-commerce websites like E-Bay, Flipkart, Amazon etc. that printing Aadhaar card on the plastic card and charging for it is a punishable offence, and the e-commerce firms are liable to pay for it and may lead to imprisonment too under the IPC (**I**ndian **P**enal **C**ode), and also under Chapter VI of Aadhaar Act (Aadhaar Act 2016) which states that "The Authority shall take all necessary measures to ensure that the information in the possession or control of the Authority, including information stored in the CIDR (**C**entral **I**dentities **D**ata **R**epository), is secured and protected against access, use or disclosure not permitted under this Act or regulations made there under, and against accidental or intentional destruction, loss or damage".

Also, according to the sources of India Times [49], UIDAI asked the Play store to take down 12 of the fraudulent applications which violated Aadhaar act 2016. UIDAI stated that the application owners without the permission of UIDAI, were giving the services like downloading Aadhaar card, generating Aadhaar status like service which led them get the access to user's sensitive data and enrolment number. It is also to be noticed that UIDAI has not authorised any of these firms to extend any Aadhaar related services. Therefore UIDAI asked the Google play store to shut down all those application. Further a warning was also given to use the Aadhaar logo in any of the application, which is again illegal according to the Aadhaar Act and Copyright Act.

# 4 Supreme Court of India's Findings

Aadhaar card is a major concern for the nation which led the Supreme Court of India interferes. The objective of this section is to present the findings of the court. On 20th November of 2012, the legislative and the state knocked the door of Supreme Court of India where the court observes the arguments against National Identification Authority of India Bill 2010 which possibly overlaps the Article 73 of Constitution of India which states "Extent of the executive power of the Union, states that, Subject to the provisions of this Constitution, the executive power of the Union shall extend to the matters with respect to which Parliament has power to make laws and to the exercise of such rights, authority and jurisdiction as are exercisable by the Government of India by virtue of any treaty or agreement". On 23$^{rd}$ September of 2013, the Supreme Court of India held by three-judge bench ordered that the Central Govt. cannot refuse to give subsidies to the person who does not possess an Aadhaar card. Therefore, court admitted that Aadhaar is voluntary but not mandatory. But in 2016, the Supreme Court of India extended the use of Aadhaar Card to MGNREGA, pension schemes, EPF (**E**mployee **P**rovident **F**und) and PMJDY though Aadhaar was first only restricted to Cooking Gas subsidies and PDS distribution system. On 7$^{th}$ February of 2017, the Supreme Court of India ordered to link mobile number with Aadhaar card as well as reminds that the government cannot make Aadhaar mandatory for welfare schemes. It is to be noticed that in January 2017-March 2017 alone the government of India has made possession of an Aadhaar card mandatory for availing over 30 central schemes. On 27$^{th}$ March 2017, the court again reiterated that government cannot make Aadhaar mandatory for welfare schemes [30]. However, the court has not stopped the government to make Aadhaar mandatory for other schemes. Recently Supreme Court of India has also started hearing on a batch of petitions challenging Section 139AA of Income Tax Act which made mandatory linking of Aadhaar with IT Returns. Senior Advocate Aravind Datar argued that the Section 139AA of Income Tax Act is contrary to the orders of the Supreme Court, and further the section violates Article 14 and 19(1)(g) of the constitution of India. A similar type of petition was also filed by the former Kerala minister and CPI leader Binoy Viswam stating that section 139 AA(1) is 'illegal and subjective' and violates Article 14 and Article 21 of Constitution of India [50].

## 5 Discussion and Conclusion

Data Privacy and Data protection are rights for every individual residing in this country and every citizen must be aware of that. Every IT technician's main focus is to implement the strongest security which no one can breach. However a Network may not be 100% secure but if one cannot make it 100% secure, there should be a legislation or system to deal with data breach cases. In India some of the Data Privacy and Protection laws are somewhat included under the sections of 'IT Act (2008)', and many are yet to be implemented and are under consideration. In Aadhaar act there are also some issues need to be understood. This section deals with such issues.

- It is to be noticed that Aadhaar is not a unique identity card, it's just a number. It does not contain any security feature like PAN card and Voter ID does. Multiple copies of Aadhaar can be downloaded from their SSUP portal. Yet, it is used as valid proof of identity at some places.
- Aadhaar is used as proof of address. UIDAI doesn't even verify the address of the applicants. Still, Aadhaar has been accepted as a proof of address in banking sector and telecom companies.
- Aadhaar cannot be used as a proof of citizenship as Clause 9 of Chapter III Aadhaar Act 2016 which states that "The Aadhaar number or the authentication thereof shall not, by itself, confer any right of, or be proof of, citizenship or domicile in respect of an Aadhaar number holder." Also, Aadhaar Act 2016, Chapter I Clause 2(v) states that "If a resident who has resided in India for more than 182 days or more than that is applicable to enrol for Aadhaar". For applying passport, Aadhaar can be used as a valid document for proof of concept and on the other hand, passport becomes a proof of citizenship which has lead to various immigrants from Nepal, Bangladesh, Bhutan etc. to get valid documents for Indian Citizenship.

- Recently, twitter's top story was MS Dhoni Aadhaar data got leaked which also depicts some flaws in Aadhaar Act which is Chapter VI Clause 4 of Aadhaar Act 2016 depicts that "No Aadhaar number or core biometric information collected or created under this Act in respect of an Aadhaar number holder shall be published, displayed or posted publicly, except for the purposes as may be specified by regulations." Also Clause 30 of IT Act 2000 states that biometric or demographic data are recognized as an 'electronic and sensitive data of an individual' and Chapter VII Clause 37 of Aadhaar Act 2016 states "Whoever, intentionally discloses, transmits, copies or otherwise disseminates any identity information collected in the course of enrolment or authentication to any person not authorised under this Act or regulations made there under or in contravention of any agreement or arrangement entered into pursuant to the provisions of this Act, shall be punishable with imprisonment for a term which may extend to three years or with a fine which may extend to ten thousand rupees or, in the case of a company, with a fine which may extend to one lakh rupees or with both".

As the major concern is the security and privacy of the data, UIDAI soon going to adopt a new encryption standard on the Aadhaar biometric devices from June 1, 2017. The new encryption standard would be added as a third layer of security. First layer is the encryption from merchants/agency side and second layer is from UIDAI itself. Third layer which is being added is implemented in the biometric device itself. UIDAI officials have informed the vendors and merchants to let their device go through the STQ (**S**tandardization **T**esting and **Q**uality) certification. Therefore, UIDAI ensured that only registered devices are allowed to make Aadhaar transaction. The objective is to tighten the security as devices are set to take biometric-based digital payments. Although the UIDAI going to take good initiatives but still there are some questions over the accuracy of biometrics, as thumb impression and iris of citizens may get changed or damaged who are involved in casual labour, and the chance of a false positive in India is 0.057%. A large portion of Indian population is involved in casual labour so it may result lakhs of false results. In fact some reports from Rajasthan and other states of India have already been received that biometrics scans are not showing a match. Therefore the usage of biometric may also lead to major security threats. However the government of India has already spent a lot of money on the Aadhaar project, and the findings of the paper entitled 'A cost-benefit analysis of Aadhaar' published by The National Institute of public finance suggest that the benefits of the Aadhaar project will surpass the costs.

## Acknowledgment

Authors thank the Editor and anonymous reviewers for their valuable comments which lead to substantial improvement on an earlier version of this manuscript. First Author expresses his sincere gratitude to the Research Mentors of Accendere Knowledge Management Services Pvt. Ltd.. Although any errors are my own and it should not tarnish the reputation of these esteemed persons.